**A Roadmap for Applying Graph Neural Networks to Numerical Data: Insights from Cementitious Materials**

Mahmuda Sharmin[1], Taihao Han[1*], Jie Huang[1], Narayanan Neithalath[2], Gaurav Sant[3] Aditya Kumar[1]

1. Missouri University of Science and Technology, Rolla, MO, USA 65409
2. School of Sustainable Engineering and the Built Environment, Arizona State University, Tempe, AZ, USA 85287
3. Civil and Environmental Engineering, University of California, Los Angeles, Los Angeles, CA, USA 90095

**<u>*Corresponding Author</u>**

Taihao Han

Assistant Research Professor, Department of Materials Science and Engineering

Missouri University of Science and Technology

248A McNutt Hall, 1400 N Bishop, Rolla, MO 65409

Email: thy3b@mst.edu; Phone: 573-341-6051; Fax: 573-341-6934

**Abstract**

Machine learning (ML) has been increasingly applied in concrete research to optimize performance and mixture design. However, one major challenge in applying ML to cementitious materials is the limited size and diversity of available databases. A promising solution is the development of multi-modal databases that integrate both numerical and graphical data. Conventional ML frameworks in cement research are typically restricted to a single data modality. Graph neural network (GNN) represents a new generation of neural architectures capable of learning from data structured as graphs—capturing relationships through irregular or topology-dependent connections rather than fixed spatial coordinates. While GNN is inherently designed for graphical data, they can be adapted to extract correlations from numerical datasets and potentially embed physical laws directly into their architecture, enabling explainable and physics-informed predictions. This work is among the first few studies to implement GNNs to design concrete, with a particular emphasis on establishing a clear and reproducible pathway for converting tabular data into graph representations using the k-nearest neighbor (K-NN) approach. Model hyperparameters and feature selection are systematically optimized to enhance prediction performance. The GNN shows performance comparable to the benchmark random forest, which has been demonstrated by many studies to yield reliable predictions for cementitious materials. Overall, this study provides a foundational roadmap for transitioning from traditional ML to advanced AI architectures. The proposed framework establishes a strong foundation for future multi-modal and physics-informed GNN models capable of capturing complex material behaviors and accelerating the design and optimization of cementitious materials.

## 1.0. Introduction

Cement is a multicomponent, complex material composed primarily of alite, belite, tricalcium aluminate, and tetracalcium aluminoferrite. Even minor variations in the proportions of these phases can profoundly affect the final performance. Moreover, the quality of raw materials—particularly the minerals sourced from different regions—introduces variability between production batches and across manufacturing sites, leading to performance inconsistencies. Although cements have been used since ancient times to the modern world, our scientific understanding of their fundamental nature remains incomplete. For example, despite extensive studies, the mechanisms governing cement hydration are still debated [1,2]. Furthermore, the relationships between cement composition and performance are highly nonlinear, adding another layer of complexity to understanding and investigating cement chemistry. Despite these challenges, researchers have attempted to develop analytical models that correlate cement compositions to their performance [3–6]. In recent years, supplementary cementitious materials (SCMs) have been widely adopted to partially replace PC in concrete, reducing both cost and environmental impact. Some SCMs (e.g. quartz and limestone) are mostly inert and primarily serve as additional nucleation sites for hydration products. Their influence on the binder performance is relatively simple. However, many others (e.g. fly ash and slag) actively react with cement. Their chemical composition and crystallinity vary significantly among sources, leading to large differences in reactivity and reaction mechanisms (e.g., pozzolanic reaction and sulfur retardation). Consequently, developing analytical models that can accurately capture the complex interactions among cement, SCMs, and other parameters (e.g. curing conditions) remains a major challenge in cement science.

To address this challenge, researchers have widely adopted machine learning (ML) as a powerful tool for predicting and optimizing performance of cementitious materials. ML provides a data-driven alternative to traditional analytical models, enabling the discovery of complex underlying correlations between all influential parameters and properties without requiring explicit mechanistic equations. Numerous studies have demonstrated the capability of ML to accurately predict mechanical properties [7–9], rheological behaviors [10–12], hydration kinetics [13–15], and durability [16–18] using mixture designs as inputs. In addition to numerical datasets, researchers have recently explored the use of microstructural [19–21] and spectral [22,23] imaging data as ML to predict concrete performance. The ability of ML to learn from image-based data has opened new pathways for linking microstructure to macroscopic behavior, providing insights that are difficult to extract through conventional statistical or analytical methods. Despite these advancements, one of the primary challenges in applying ML to cementitious materials is the limited size of the database. Acquiring a single experimental data point can require extensive laboratory work—often involving multiple stages of material preparation, curing, testing, and data processing—which makes assembling datasets with thousands of records time-consuming and costly. Consequently, many existing ML models are trained on relatively small, homogeneous datasets, which restricts their generalizability and robustness. A promising strategy to overcome this limitation is to develop multi-modal databases that integrate both numerical data and image-based data. Such databases can be constructed by systematically collecting and harmonizing data from published literature, ensuring wide coverage of materials, compositions, and testing conditions. Integrating heterogeneous data sources not only increases the database size and variability but also provides a more holistic representation of the cementitious system. Another advantage of developing multi-modal databases is that numerical and graphical data provide complementary perspectives on the same property. Numerical data capture quantitative relationships among mixture designs and processing conditions, while graphical data reflect underlying microstructure and morphology. Integrating these distinct yet related forms of information can enhance the generalizability and reliability of ML models. However, current ML frameworks used in cement research are typically designed to process a single data modality—either numerical or image-based inputs—limiting their ability to fully exploit the complementary information embedded across different data types. Therefore, there is an urgent need for developing advanced, multi-modal ML architectures capable of simultaneously learning from numerical and graphical data.

Graph neural network (GNN) represents a new class of neural networks specifically designed to extract and learn features from data represented in graph structures [24–26]. Unlike traditional convolutional neural networks that operate in Euclidean space, where data points (or neurons) are arranged on regular grids with uniform spacing—such as pixels in an image—GNN functions in non-Euclidean space, where the relationships between data elements are defined by edges connecting nodes rather than by spatial proximity. This allows GNN to discover underlying correlations where interactions occur through complex, irregular, or topology-dependent connections rather than fixed spatial coordinates. In essence, GNN generalizes the concept of convolution from Euclidean domains to irregular graph domains, enabling feature extraction from data structures such as molecular networks, social networks, or material microstructures. Each node in a graph represents an entity (e.g., an atom, mineral phase, or voxel of a microstructure), and edges define their physical, chemical, or spatial relationships. Through iterative message-passing operations, each node updates its representation by aggregating information from its neighboring nodes and edges, allowing the network to learn latent correlations and hierarchical dependencies within the data. This architecture makes GNN particularly powerful for predicting materials properties through learning physical and chemical information from molecular structure, microstructure, and molecular dynamics [27–30]. GNN model has been applied to optimize the concrete structures [31,32] and predict energy consumption of cement manufacturing [33]. However, GNN model has never been applied to correlate composition with performance of cementitious materials. It has big potential to translate microstructure of concrete into graph representations, where nodes denote individual phases, pores, or features, and edges encode their geometric or physicochemical interactions. By learning from such structured representations, GNN can effectively capture local connectivity patterns, long-range dependencies, and complex topological relationships that are often overlooked by conventional ML models.

Recent studies have demonstrated that GNN is not limited to graphical data—they can also be effectively adapted to learn from tabular databases. Several researchers have proposed innovative strategies to convert tabular data into graph structures, allowing GNNs to process and interpret numerical features in a relational context. For example, Li et al. [34] introduced multiple graph-construction approaches (i.e., data as nodes, features as nodes, and data and features as nodes) to transform tabular datasets into graph representations. Hong et al. [35] developed a GNN-based architecture designed for information exchange, in which each node aggregates information from multiple numerical features. The most informative node embeddings are then passed to subsequent neural network layers to predict melting temperature of materials. Guo et al. [36] constructed a multiplex graph that models multifaceted relationships among samples, and by designing a multiplex GNN capable of learning enhanced, high-dimensional feature representations through cross-layer interactions. However, a clear pathway for applying GNNs to numerical data remains underexplored, particularly in materials science and cement chemistry.

As discussed earlier, improving the data quality and diversity for cementitious materials requires building multi-modal databases that integrate both numerical and graphical data. In this context, GNN is a promising solution for learning complex correlations and generating reliable predictions across such heterogeneous data domains. An additional advantage of GNN is their inherent flexibility in defining nodes, edges, and connectivity rules, which enables the explicit incorporation of physical laws and domain knowledge into the model. For instance, nodes can be defined at different hierarchical levels, and edges can be selectively established based on physically meaningful interactions. Moreover, customized activation functions can be designed for specific edges to reflect known correlations, ensuring that model predictions remain physically consistent. This ability to embed physics directly within the network architecture positions GNNs as a transformative tool for developing next-generation, explainable, and physics-informed models for cementitious materials.

This work is among the first few studies to implement GNNs to learn composition-performance correlations and design concrete, emphasizing the exploration of strategies for implementing GNN on tabular databases. Although the current work focuses exclusively on tabular data, it marks an important step toward developing multi-modal (i.e., numerical and graphical) frameworks. To transform tabular inputs into a graph structure, a k-nearest neighbor (K-NN) model is employed to map tabular data-records into graphical space, thereby defining the nodes and edges of the GNN. The optimal number of neighbors of K-NN is systematically determined. Furthermore, feature importance analysis is conducted to refine input selection and evaluate how feature modification influences GNN performance. The prediction performance of GNN model is compared with benchmark random forest (RF) model. Overall, this study establishes a foundational framework for transitioning from traditional ML to advanced AI models, leading to produce accurate predictions and construct physically meaningful architecture of GNN for cementitious materials.

## 2.0. Graph Neural Network (GNN)

GNN model is a ML model designed to learn correlations from the graph-structured data. GNN model can directly handle practically useful graph types, such as cyclic, acyclic, directed, and undirected graphs. Unlike traditional neural networks as multilayer neural networks, convolutional neural networks, or recurrent neural networks that operate on fixed-size vector or grid inputs [37], GNNs were designed to learn from non-Euclidean, graph-structured data through a mechanism known as message passing [38–40]. This mechanism is the core process of GNNs that allows them to flow the information through connected nodes, which allows the network to learn both local dependencies (from immediate neighbors) and global graph topology through multiple layers of aggregation.

In a graph in GNN, entities are represented as nodes ($V$), while relationships or similarities between them are represented as edges ($E$). This graph representation allows the model to leverage structural similarities between samples to improve learning accuracy. Before message passing begins, each node $i \in V$ is initialized with a d-dimensional feature vector that represents its input attributes, as

$$x_i = h_i^0 \in \mathbb{R}^d \quad \text{Eq. 1}$$

Here, $x_i$ denotes the input feature vector of node $i$ and $h_i^{(0)}$ represents its initial state at the input layer of the GNN. During the message passing process, each node $i$ updates its hidden information or features $h_i^{(k)}$ at layer $k$ by aggregating information from its neighboring nodes $j \in N(i)$. It is an iterative process through which each node aggregates information from its neighbors and updates itself by combining its own information. This message passing occurs in two steps: aggregation and combination, expressing as

$$a_i^{(k)} = AGGREGATE^{(K)}\{ h_j^{(k-1)} : j \in N(i)\} \quad \text{Eq. 2}$$

$$h_i^{(k)} = \sigma\ (W^{(k)} \cdot COMBINE^{(K)}\{ h_i^{(k-1)}, a_i^{(k)}\}) \quad \text{Eq. 3}$$

Here, $N(i)$ is set of neighboring nodes of $i$ and $h_i^{(k-1)}$ is the node's features in the previous iteration $k-1$. For every node, the *AGGREGATE* function gathers the features $h_j^{(k-1)}$ from all its neighboring nodes $j \in N(i)$ and produce an intermediate aggregated message $a_i^{(k)}$. The *COMBINE* function then merges this aggregated message with the node's own previous representation $h_i^{(k-1)}$ to produce the updated node feature representation $h_i^{(k)}$. Then the update of each node is multiplied by a learnable weight matrix $W^{(k)}$ to transform the aggregated information into an updated node representation and a non-linear activation function $\sigma$ is applied to introduce nonlinearity into the model. In this study, the rectified linear unit (*ReLU*) activation function was used which has been widely adopted in modern GNN architectures such as GCN and GraphSAGE [41,42]. *ReLU* efficiently mitigates the vanishing-gradient problem often encountered in deep neural networks, while maintaining computational simplicity and sparse activations. These features accelerate convergence during training and improve generalization performance.

When given the graph structure and node information, depending on the type of prediction task, GNNs can operate at three levels: node-level, edge-level, and graph-level [40,43].

- In node-level tasks, the model estimates properties for individual nodes, such as classification or regression, based on the node's features and those of its neighbors.
- In edge-level tasks, the model predicts relationships or interactions between two connected nodes, including their existence, type, or strength edge.
- In graph-level tasks, embeddings of all nodes are combined using mean, sum, or attention pooling to create a single graph representation, which is then employed to predict a global property like molecular activity or material class.

Different GNN architectures have developed to achieve message-passing, utilizing various aggregation and update schemes. These architectures share the same conceptual backbone which is the message-passing framework but differ in how neighboring information is aggregated, normalized, or weighted. Collectively, they enable GNNs to learn hierarchical and relational representations from complex, non-Euclidean data structures.

- Graph Convolutional Network (GCN): GCN performs feature propagation using a normalized adjacency matrix. Each node updates its representation by combining its own features with those of its neighbors, then applies a linear transformation and nonlinear activation [41]. GCN works best for homophilous graphs, where connected nodes share similar properties.
- GraphSAGE: GraphSAGE learns node embeddings by sampling a fixed number of neighbors, aggregating their features, and concatenating them with the node's own feature before applying a linear transformation. This allows inductive learning on unseen nodes or graphs [42]. GraphSAGE is well-suited for large-scale or dynamic graphs due to its sampling strategy.
- Graph Attention Network (GAT): GAT incorporates self-attention to assign different importance weights to neighboring nodes during aggregation, enabling adaptive feature propagation [44]. GAT is effective for heterogeneous graphs, where the importance of neighbors varies.
- Message Passing Neural Network (MPNN): MPNN defines a general message-passing framework consisting of a message phase and an update phase. Edge features can be integrated seamlessly, making them ideal for molecular and physical systems [39]. MPNN is well-suited for graphs with many edges, such as molecules, where bond details are crucial.

In this study, GraphSAGE architecture in conjunction with mean aggregation was implemented using the PyTorch Geometric library to perform supervised node-level regression. For the concatenation part, mean aggregation was used. Each node in the constructed graph represents a concrete mixture design and the output node is the compressive strength. To define the relationships (edges in GNN) between samples, a K-NN model [45,46] was applied to the normalized feature space to create edges between compositionally similar nodes. This strategy transforms tabular data into a graph structure, allowing the GNN to propagate information across similar nodes and learn relational dependencies. The resulting graph was made undirected by adding reciprocal edges, ensuring bidirectional information exchange between connected nodes.

Using K-NN can be challenging because it requires tuning hyperparameters. The key hyperparameters include the number of neighbors, the distance function, and the weighting function [46]. In this paper, a sensitivity study was performed over a wide range of K values (2 to 30) to examine the influence of graph connectivity on model performance. Lower K-NN values may cause sparse connections and underfitting, while higher values may lead to over-smoothing by incorporating excessive neighbor information. The K-NN construction allows the model to interpret each data-records in the context of its most similar neighbors rather than in isolation. Euclidean function [47,48] was utilized to calculate distance between data-records.

The GraphSAGE layer was selected because it performs inductive learning, which means that it can generalize to unseen data or new nodes without requiring retraining on the entire graph [42,44]. Unlike spectral approaches such as the GCN [41,43], which depend on a fixed adjacency matrix, GraphSAGE learns aggregation functions that sample and combine information from a node's neighborhood. This approach is particularly suitable for datasets like concrete mixtures, where each node (sample) is independent but shares compositional relationships with others.

The network architecture of this paper consisted of three GraphSAGE convolutional layers with 128 hidden neurons, each followed by batch normalization, ReLU activation, and dropout (0.25) to prevent overfitting. The model parameters were optimized using the Adam optimizer with a mean squared error (MSE) loss function, consistent with other regression-based GNN studies [41, 42]. A fixed random seed (42) was used to ensure reproducibility, so that model initialization, data splits, and training outcomes remained consistent across multiple runs. Through multiple layers of message passing and nonlinear transformation, the network learned to map node features and neighborhood information to the predicted compressive strength.

To illustrate the workflow of the implemented GNN model, **Figure 1** presents the complete pipeline used for model construction, training, and evaluation. The process begins with input data normalization and graph construction using the K-NN, followed by training a GraphSAGE model that performs message passing to aggregate information from neighboring nodes. During each forward pass, node features are propagated and combined to form updated node representations, while the training loss is computed and minimized through backpropagation and weight updates like traditional neural network. Model performance on the validation set is continuously monitored, and the model is saved whenever improvement is detected. If no improvement occurs for 30 consecutive epochs, training is terminated early to prevent overfitting. Finally, the best-performing model is reloaded for testing and evaluation on unseen data, and then the predictions are converted back to their original physical units (MPa).

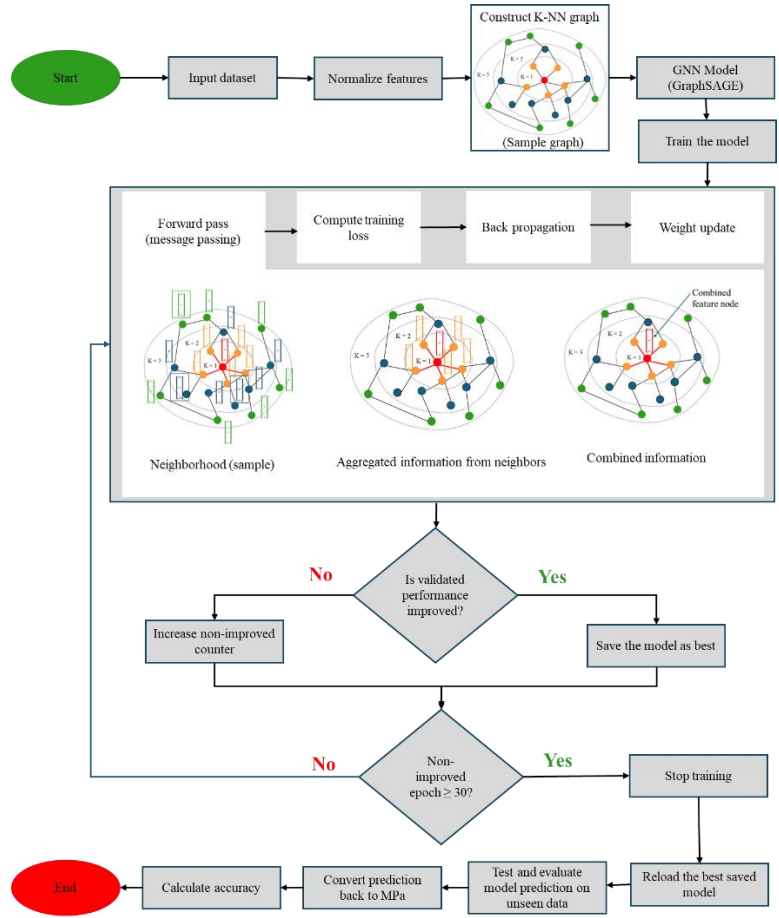

**Figure 1**: Workflow of the GNN model for predicting concrete compressive strength from mixture designs, encompassing data preprocessing, model implementation, and performance evaluation.

## 3.0. Database Collection, Data Processing, and Random Forest (RF) Model

The database used in this study is the concrete compressive strength dataset from the UCI Machine Learning Repository [49]. This well-established database has been widely utilized for evaluating performance of various ML models. Because the database is relatively straightforward and well-structured, it allows readers to focus on understanding the implementation and performance of GNN model. Moreover, using this benchmark database facilitates direct comparison between the GNN and conventional ML models (i.e., RF). Employing such a standard benchmark is crucial for assessing the advantages and limitations of the proposed GNN framework.

The concrete database comprises 1030 data-records, each containing eight input variables and one output variable, with no missing values in the database. The input variables are cement content (kg/m$^3$); fly ash content (kg/m$^3$); slag content (kg/m$^3$); superplasticizer (kg/m$^3$); fine aggregate content (kg/m$^3$); coarse aggregate content (kg/m$^3$); water content (kg/m$^3$); and age (day). The output is time-dependent compressive strength (MPa). The statistical parameters pertaining the data range and distribution can be found in our previous study [7]. Before building the graph and inputting data into the GNN model, data preprocessing steps were performed to maintain numerical consistency and make features comparable. Normalization [50–53] is applied to the database, which ensures input variables ranging from 0 to 1. It allows that every feature contributes equally to graph generation and message passing, reducing the bias from the database.

To perform the training and validation of GNN model, the compressive strength database was randomly split into three datasets: (1) 70% of data-records for training dataset; (2) 15% of data-records for validation dataset; and (3) the remaining 15% of data-records for testing dataset. To evaluate the performance of GNN model, four statistical metrics are utilized: coefficient of determination ($R^2$), mean absolute error ($MAE$), root mean squared error ($RMSE$), and mean absolute percentage error ($MAPE$). These metrics quantify the differences between the predicted and measured compressive strength.

RF model was employed in this study to enable direct performance comparison with the GNN model. A brief overview of RF is provided below, while further details can be found in our previous studies [14,54,55]. RF is an ensemble learning method based on classification and regression trees, designed to enhance prediction performance through two components: bootstrap aggregation (bagging) and randomization [56,57]. Specifically, RF constructs multiple decision trees using randomly generated subsets of the training data, where each subset is created by sampling data records from the parent dataset with replacement. At each node within a tree, RF introduces an additional layer of randomness by selecting a random subset of input variables for node splitting. RF allows trees to grow to their full depth, thereby promoting greater diversity among individual trees. During prediction, RF aggregates the outputs from all trees—typically by averaging in regression tasks—to produce the final output.

## 4.0. Results and Discussion

In this study, two types of GNN tasks are implemented: node-level and graph-level regressions. The primary focus is on node-level regression, as it aligns more closely with the nature of the database and effectively preserves mixture design information during training. For comparison, graph-level regression is also conducted using the same dataset. Feature selection plays a crucial role in determining the prediction performance of ML. In cementitious materials, features can be represented in multiple ways with the same meaning—for example, water and cement content can be treated as independent inputs or combined as the water-to-binder ratio. To reduce computational demand and improve model efficiency, it is often desirable to minimize the number of input variables. Therefore, the first step of this study involves evaluating different feature combinations to assess their impact on prediction accuracy.

The RF model was employed to evaluate the influence of each input variable on the prediction performance using the Gini importance score. Details on how variable importance is calculated using the RF model can be found in our previous study [58]. **Figure 2** demonstrates the influences of each variable on prediction accuracy of the compressive strength. As expected, cement content, water content, superplasticizer dosage, and curing age exert the most significant influence on compressive strength. Cement serves as the primary source of strength through hydration reactions; water content governs porosity, which is directly correlated with strength; and an appropriate amount of superplasticizer enhances packing density and reduces entrapped air. Moreover, compressive strength typically increases with age due to the continued formation of hydration products. Following these dominant variables, the effects of blast furnace slag, fine aggregate, coarse aggregate, and fly ash contents were observed in less influence on the compressive strength.

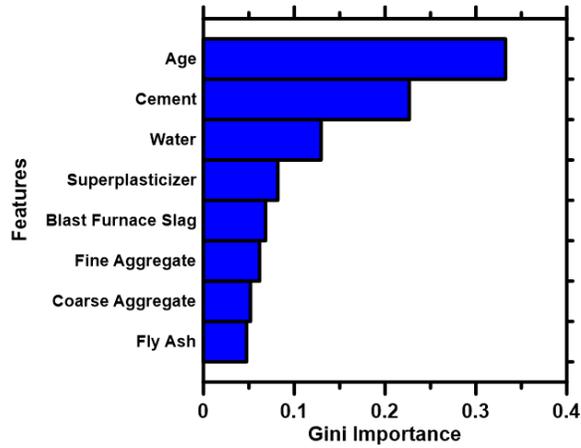

**Figure 2**: Feature importance of input variables affecting concrete compressive strength as determined by RF model. Variables are ranked in descending order according to their importance.

Based on the feature importance analysis and domain knowledge of cement chemistry, the input variables for the GNN model were reorganized into five feature groups. These groups involve both the removal of less influential variables and the combination of correlated inputs into single representative features. **Table 1** summarizes the five feature groups and their corresponding variables. The main objective of this approach is to evaluate whether the GNN model can maintain reliable predictions while using a reduced number of input variables. The original dataset is retained by feature group A. Feature group B includes the top 4 important input variables shown in **Figure 2**. To mitigate multicollinearity in features, several input variables are unified into single input variable based on cement chemistry, resulting in the formulation of feature group C. Feature group D is designed based on combining similar materials into one group. Based group D, additional combining on similar materials is performed on feature group E.

| Table 1: Definition of five feature groups used in the GNN models, aimed at reducing the number of input variables and consolidating multiple correlated features into unified representations. | | |
|---|---|---|
| **Feature group** | **No. of features** | **Input features** |
| A | 8 | 1. Cement; 2. Slag; 3. Fly Ash; 4. Water; 5. Superplasticizer; 6. Coarse Aggregate; 7. Fine Aggregate; 8. Age |
| B | 4 | 1. Cement; 2. Water; 3. Superplasticizer; 4. Age |
| C | 7 | 1. Binder = Cement + Slag + Fly Ash; 2. Aggregate = Coarse + Fine Aggregate; 3. Fluidity = Water + Superplasticizer; 4. w/b = Water/Binder; 5. Agg/b = Aggregate/Binder; 6. sp/b = Superplasticizer/Binder; 7. age |
| D | 7 | 1. Cement; 2. Slag; 3. Fly Ash; 4. Water; 5. Superplasticizer; 6. Aggregate = Coarse + Fine Aggregate; 7. Age |
| E | 6 | 1. Cement; 2. SCM = Slag + Fly Ash; 3. Water; 4. Superplasticizer; 5. Aggregate = Coarse + Fine Aggregate; 6. Age |

The GNN model is implemented for all five feature groups. To achieve optimal performance, the hyperparameters are carefully fine-tuned. In this study, the most critical hyperparameter is the number of nearest neighbors in the K-NN, which defines how many edges that each node has. **Figure 3** shows the relationship between the number of neighbors and model accuracy for each feature group, while **Table 2** summarizes the statistical performance metrics corresponding to the optimal number of neighbors for each group.

**Figure 3** demonstrates that the prediction performance was highly sensitive to the choice of the neighborhood size. When neighborhood size is small, the resulting graph is sparse—many nodes have few or no connections, which limits message passing and increased variance in predictions. Conversely, when neighborhood size becomes large (K > 8), the graph turns overly dense, with most nodes connect to nearly all others. This excessive connectivity lead to over-smoothing, a common issue in message-passing GNNs such as GraphSAGE, where node representations become overly similar due to excessive information aggregation from neighboring nodes [42,43,59]. Across most feature groups, the optimal neighborhood size ranges 3 to 5 of 3–5. Because feature groups A and E shows relatively higher accuracy, prompting further investigation into denser neighborhood configurations (as shown in **Figures 3c** and **3d**). Increasing neighborhood size beyond this range, however, results in a noticeable decline in prediction accuracy. The optimal choice of neighborhood size appears to depend on two key factors: the database size and the number of input variables. Larger datasets or those with more input features may require higher neighborhood size to effectively capture representative relationships. For instance, feature group B, which included only four input variables, achieved optimal performance with neighborhood size at 2. Future studies should further verify this relationship to establish a more generalized guideline for selecting neighborhood size in GNN models for cementitious materials.

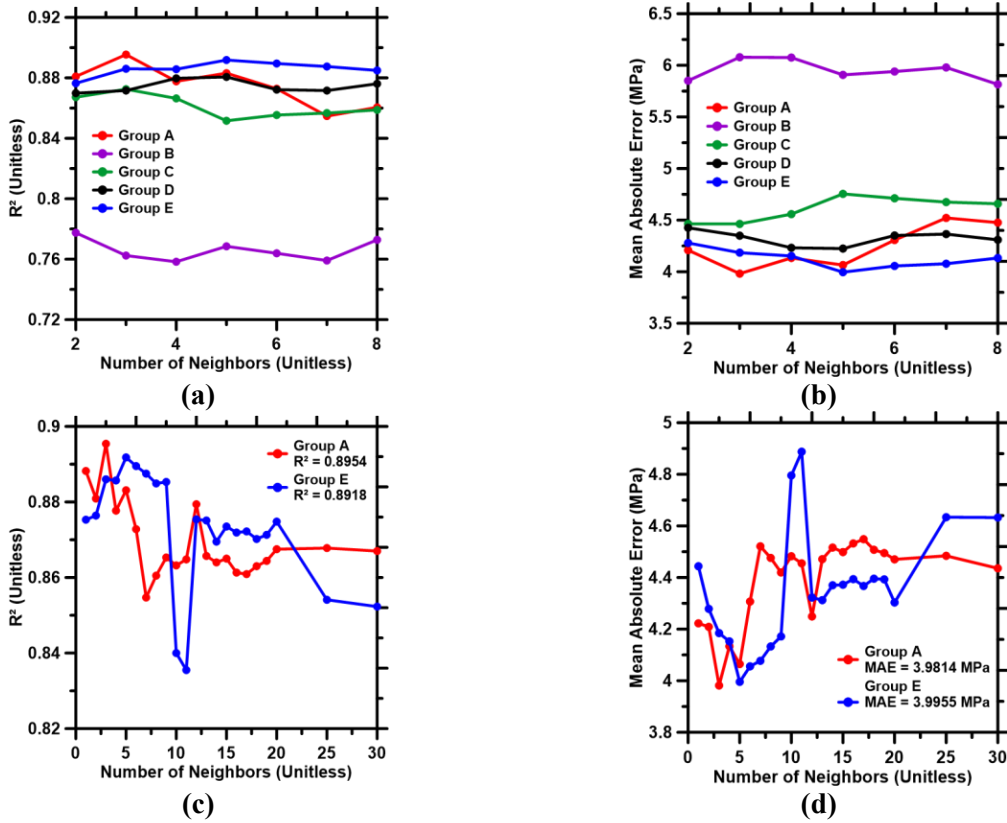

**Figure 3**: Prediction performance of the GNN model for different feature groups as a function of the number of neighbors. (a) and (b) show the model performance for all feature groups at smaller neighborhood sizes, while (c) and (d) illustrate the performance of feature groups A and E under larger neighborhood sizes.

| Table 2: Prediction performance of the GNN model with different feature group and task type and their optimal number of neighbors. | | | | | |
| --- | --- | --- | --- | --- | --- |
| **GNN type and feature group** | **No. of neighbors** | $R^2$ **(unitless)** | *MAE* **(MPa)** | *MAPE* **(%)** | *RMSE* **(MPa)** |
| Node level - A | 3 | 0.8954 | 3.98 | 14.48% | 5.07 |
| Node level - B | 2 | 0.7775 | 5.85 | 20.40% | 7.39 |
| Node level - C | 3 | 0.8725 | 4.46 | 16.54% | 5.59 |
| Node level - D | 5 | 0.8806 | 4.22 | 15.29% | 5.41 |
| Node level - E | 5 | 0.8918 | 3.99 | 15.08% | 5.15 |
| Graph level - A | 3 | 0.8138 | 5.29 | 19.48% | 6.98 |

In **Table 2**, it is clear that feature group A with a neighborhood size of 3 achieves the best prediction performance, yielding an $R^2$ of 0.8954 and an *MAE* of 3.98 MPa. Such strong performance is expected since group A contains all essential mixture design parameters that directly influence compressive strength. In contrast, group B, which included only four key features, demonstrates the lowest prediction accuracy, regardless of neighborhood size. This confirms that the absence of key mixture design information constrains model capability of learning hidden input-output correlations. Groups C, D, and E demonstrate good prediction performance despite using fewer input variables. Their success stems from combining multiple original features into composite inputs, allowing the model to retain critical information while reducing computational cost. Groups A and E, which have superior performance, will be used for the following investigations. As shown in the table, the graph-level task—performing with feature group A—underperforms relative to the node-level model. This decline is attributed to the global pooling operation in graph-level tasks, which averages information across the entire graph and consequently loses sample-specific detail. Further analysis of graph-level task is discussed in the following sections.

To ensure optimal model performance, the number of training epochs is another critical hyperparameter that requires careful fine-tuning. In a neural network, the epoch refers to the number of times the entire training dataset is passed through the model during training. Each epoch updates the internal weight functions based on the *MSE* between predictions and actual values, gradually moving the model toward the global minimum of the loss function. An insufficient number of epochs may lead to underfitting, while excessive epochs can cause overfitting and reduce generalization capability.

**Figure 4** presents the variation in *RMSE* for the training, validation, and testing datasets as the model was trained over multiple epochs. The blue, orange, and green curves correspond to the training, validation, and testing datasets, respectively. Each plot terminates at a different epoch because an early-stopping condition was applied—training was halted if the validation loss failed to improve by at least $1\times10^{-4}$ over 30 consecutive epochs. For both node-level and graph-level regressions, the GNN models used neighborhood size of 3 for group A and neighborhood size of 5 for group E, as these configurations yielded the most accurate predictions for their respective datasets.

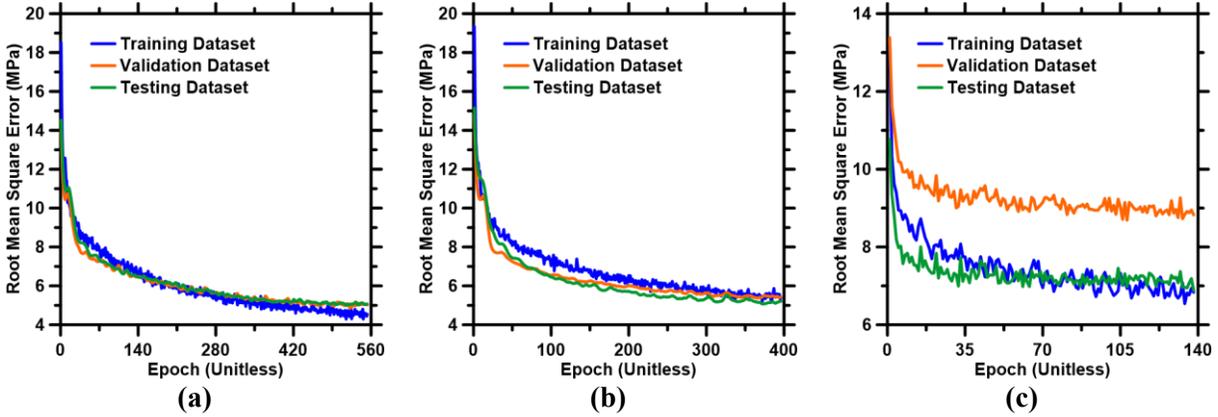

**Figure 4:** Prediction accuracy of the GNN model as a function of training epochs for: (a) node-level task using feature group A with a neighborhood size of 3; (b) node-level task using group E with a neighborhood size of 5; and (c) graph-level task using group A with a neighborhood size of 3.

In **Figure 4a**, the *RMSE* decreases rapidly during the first 100 epochs, indicating that the model quickly captured the nonlinear relationships between mixture design and compressive strength. As training progressed, the curves gradually flattened, signifying convergence and stable model performance. A steady decline is observed up to approximately 280 epochs, after which all three curves stabilized—at around 4.69 MPa for the training dataset, 5.08 MPa for the validation dataset, and 5.07 MPa for the testing dataset at epoch 523. The close agreement among the three datasets suggests that the model achieved good generalization without overfitting. The lowest validation loss occurs at epoch 523, and training is stopped at epoch 553 according to the early-stopping criterion. A similar pattern is observed in **Figure 4b**, where the *RMSE* curves flatten earlier, around 200 epochs, and stabilize at approximately 5.41 MPa for training, 5.38 MPa for validation, and 5.15 MPa for testing at epoch 366, with training terminate at epoch 396. The faster convergence in this case reflects the smaller number of input variables in feature group E compared to group A, allowing the model to learn correlations more efficiently while reducing computational demand.

In **Figure 4c**, the graph-level GNN reaches convergence much earlier than the node-level models, stabilizing by approximately epoch 108 with noticeably higher *RMSE* values—6.93 MPa for the training dataset, 8.69 MPa for the validation dataset, and 6.98 MPa for the testing dataset. This rapid stabilization, combined with higher overall errors, suggests a structural mismatch between the graph formulation and the underlying problem definition. In this study, each data point represents an independent concrete mixture design and its compressive strength. However, when all samples were combined into a single global graph and processed through a global pooling operation, the model effectively compressed multiple, distinct mixtures into a single graph-level representation per forward pass. This pooling step introduces a representation bottleneck, where diverse node-level features are aggregated into a fixed-size embedding, inevitably causing information loss and reducing the effective number of independent training samples. Such behavior aligns with the well-documented "over-squashing" and "pooling bottleneck" phenomena in GNNs, where excessive message aggregation across loosely related or independent nodes restricts model expressivity and degrades predictive accuracy [60–62]. As a result, the graph-level model shows signs of overfitting, reflected by the high prediction error for the validation dataset. This highlights the limitation of applying global pooling to datasets where each record represents an independent observation rather than interdependent entities.

While the GNN model demonstrates reliable prediction performance based on error metrics, it remains uncertain whether it can match the capability of well-established ML models. To address this, the same compressive strength database is also applied to RF model—a widely adopted model for predicting concrete properties. The RF model is trained using the same dataset and partitioning strategy as the GNN model. Hyperparameters for the RF model is optimized through 10-fold cross-validation, with the remaining 30% of the data reserved for testing. For a direct, one-to-one comparison, the GNN model's prediction results includes both the validation and testing datasets, corresponding to the same 30% of data used as the RF testing dataset. Since the validation dataset is excluded from training, it remains a blind dataset for the GNN, even though it is utilized for hyperparameter tuning. This approach ensures a fair comparison between the two models in terms of prediction accuracy and generalization capability.

**Figure 5** compares the prediction performance on compressive strengths obtained from the GNN and RF models. The node-level GNN models shows good performance, where $R^2$ = 0.8992 for feature group A and $R^2$ = 0.8879 for group E). In the node-level models, the data points are closely aligned with the diagonal dashed line, indicating that the GNN effectively captured the nonlinear relationships among mixture design and strength. In contrast, the graph-level model shows more scattered data points ($R^2$ = 0.8039), reflecting weak prediction performance and loss of sample-specific information during global pooling. When compared with the RF benchmark, the RF achieves the highest overall prediction accuracy ($R^2$ = 0.9016). Despite this slight difference, the GNN demonstrates comparable prediction performance. Unlike conventional feature-based models such as RF, GNN leverages message passing to learn higher-order dependencies and inter-feature interactions, allowing it to generalize better to blind data. Moreover, the graph topology inherently encodes material similarity, providing improved physical interpretability [42]. Despite these advantages, GNN has two main drawbacks. First, its training requires significantly higher computational resources and memory due to repeated sparse-dense matrix operations and irregular data access during message passing [63]. Second, GNN is sensitive to graph construction parameters, such as the number of neighbors, which can strongly influence model stability and performance [64]. While the RF model provides a fast and reliable baseline with excellent accuracy and minimal computational cost, the GNN provides greater flexibility, interpretability, and scalability for large or relational materials datasets which are the key attributes for advancing data-driven materials design, even if higher computational resources are required. These features position GNN as promising alternatives to traditional ML models for cementitious materials.

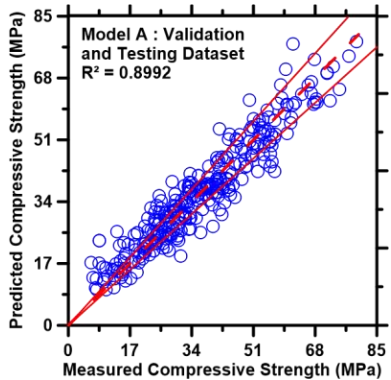

(a)

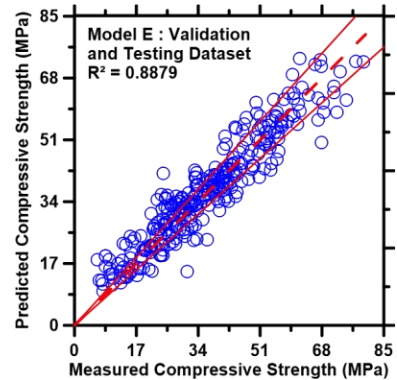

(b)

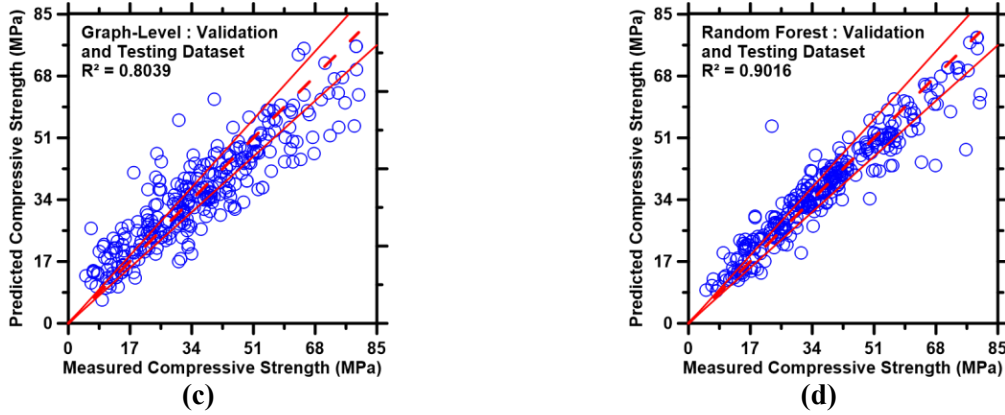

**Figure 5**: Predictions of compressive strength of concrete produced by GNN models ((a) Node-level: data model A (K-NN=3), (b) Node-level: data model E(K-NN=5), (c) Graph-level GNN(K-NN=3)); and (d) RF. The red dashed line shows the ideal predictions, and the red solid lines indicate a $\pm$ 10% error boundaries.

## 5.0. Conclusions

This study implements GNN to tabular database for cementitious materials, establishing a clear and reproducible pathway for transforming conventional mixture design inputs into graph-structured representations. By leveraging a K-NN strategy for graph construction, combined with feature-selection schemes informed by domain knowledge, the study demonstrates that GNNs can effectively learn the nonlinear and multi-dimensional relationships governing compressive strength of concrete.

In this study, each data-record was treated as a node to construct a node-level GNN model. Based on the similarity between data points, edges were established using the K-NN, effectively transforming the tabular dataset into a graph-structured representation. This approach provides a clear and reproducible pathway for applying GNN models to numerical data. Moreover, feature selection was also systematically evaluated to balance computational efficiency and predictive accuracy. The results show that models incorporating the complete mixture design information yielded the highest prediction accuracy. However, consolidating several correlated input variables into a single representative feature significantly reduced training time while maintaining comparable accuracy. In contrast, the graph-level GNN demonstrated lower prediction accuracy due to information loss during global pooling, emphasizing that node-level architectures are more suitable for datasets where each record represents an independent mixture design.

Overall, the GNN model achieved performance comparable to that of traditional ML model, demonstrating its strong potential for advancing data-driven modeling in cementitious materials.

GNN provides unique advantages for advancing materials informatics. Their flexible graph topology allows explicit incorporation of physics laws, hierarchical material descriptors, and multi-scale relationships—capabilities that are unattainable in conventional machine-learning models. These attributes position GNN as a promising foundation for the next generation of multi-modal, physics-informed AI model. Although current limitations remain—particularly sensitivity to graph construction and substantial computational demand—the demonstrated accuracy and interpretability underscore the transformative potential of GNNs for cementitious materials.

Overall, this work establishes a foundational roadmap toward advanced AI architectures capable of unifying numerical, microstructural, and physicochemical information within a single learning framework. The methodology and insights developed here provide a critical step toward scalable, explainable, and physics-informed GNN models that can enhance concrete design, accelerate materials discovery, and unlock deeper scientific understanding of cementitious systems.

**Acknowledgement**

The authors acknowledge financial support from the Kummer Institute (Missouri S&T) and the National Science Foundation (NSF-DMR: 2034856 and 2228782).

**Data Availability**

The database used in this study was downloaded from UC Irvine Machine Learning Repository.

**Author Contributions**

**Mahmuda Sharmin:** Investigation; Formal analysis; Writing – original draft; and Methodology

**Taihao Han:** Conceptualization; Investigation; Writing – original draft

**Gaurav Sant:** Supervision; Writing – review and editing; and Funding acquisition

**Narayanan Neithalath:** Supervision; Writing – review and editing; and Funding acquisition

**Aditya Kumar:** Conceptualization; Supervision; Writing – review and editing; and Funding acquisition